\documentclass{article}
\usepackage{graphics}
\pdfoutput=1
\usepackage{amssymb,amsfonts,amsmath}
\usepackage[serbian,english]{babel}
\usepackage{multirow}
\usepackage{rotating}
\usepackage{fullpage}
\usepackage[utf8]{inputenc}

\newcommand\institute[1]{\newcommand\theinstitute{#1}}

\title{The baseline for response latency distributions}
\author{Ferm\'{\i}n MOSCOSO DEL PRADO MART\'IN}
\institute{Laboratoire de Psychologie Cognitive ({UMR}-6146), Centre National de la Recherche Scientifique \& Aix--Marseille Universit\'e I, Marseille, France}

\begin{document}
\maketitle

\begin{abstract}
Response latency -- the time taken to initiate or complete an action or task -- is one of the principal measures used to investigate the mechanisms subserving human and animal cognitive processes. The right tails of response latency distributions have received little attention in experimental psychology. This is because such very long latencies have traditionally been considered irrelevant for psychological processes, instead, they are expected to reflect `contingent' neural events unrelated to the experimental question. Most current theories predict the right tail of response latency distributions to decrease exponentially \cite{Laming:1968,Ratcliff:1978}. In consequence, current standard practice recommends discarding very long response latencies as `outliers' \cite{Luce:1986,Ratcliff:1993}. Here, I show that the right tails of response latency distributions always follow a power-law with a slope of exactly two. This entails that the very late responses cannot be considered outliers. Rather they provide crucial information that falsifies most current theories of cognitive processing with respect to their exponential tail predictions. This exponent constitutes a fundamental constant of the cognitive system that groups behavioral measures with a variety of physical phenomena.
\end{abstract}

A pervading assumption in the literature is that Response Latencies ({RL}s) follow a distribution whose right tail decreases exponentially \cite{Laming:1968,Ratcliff:1978,Luce:1986}. {RL}s are ultimately by-products of the workings of the brain, and further, of the firing patterns of heavily \emph{interconnected} neural assemblies. From this perspective, exponential tails would be a rather surprising outcome for {RL} distributions \cite{Holden:VanOrden:Turvey:2009}. They would imply that the {RL}s were generated by a Poisson process, that is, they would be \emph{independent} events, despite the interconnections between the neurons that generated them. 

More in line with the probably correlated origins of behavioral events, two recent theories have predicted that the right tails of {RL}s distributions should follow a power-law \cite{Holden:VanOrden:Turvey:2009,Moscoso:2009}. This is to say that for all times $t$ greater than a certain $t_{\mathrm{min}}$, their probability density function should be that of a Pareto distribution:
\begin{equation}
p(t) = \frac{\alpha - 1}{t_{\mathrm{min}}}\left(\frac{t}{t_{\mathrm{min}}}\right)^{-\alpha},\, \alpha > 1,\, t \geq t_{\mathrm{min}},
\label{eq:dpareto}
\end{equation}
where $\alpha$ is referred to as the \emph{scaling parameter}, and it corresponds to the slope of the straight line that is formed by the density function when plotted on log-log scale. A more precise theoretical proposal \cite{Moscoso:2009} is that {RL}s arise as the result of the ratio of two correlated normal variables: The excitability of the response effector, and the strength of the signal that excited it. Therefore the distribution of {RL}s should follow a normal ratio distribution ({NRD} \cite{Fieller:1932}). This has the further implication that the power-law right tail should have a value of the scaling parameter of exactly two \cite{Jan:etal:1999,Sornette:2001}. Such a precise tail behavior would hold irrespective of the properties of the task. It would constitute a complete description of the {RL} distribution in the far right tail, in the strong sense of having zero degrees of freedom. The scaling parameter value would therefore represent a fundamental constant of the cognitive system. Furthermore, it would group {RL}s with other well-known natural systems with identical properties, such as Ising models of ferromagnetic materials close to their critical temperature \cite{Jan:etal:1999} or the intervening times between major earthquakes \cite{Mega:etal:2004}.

Obtaining estimates of the distributions in the far right tail requires very large numbers of ideally untruncated  responses. I analyzed six large-scale databases of human responses across experimental tasks and modalities, and at different time ranges. The datasets included ocular fixation and blink durations during reading (\emph{The Dundee Corpus} \cite{Kennedy:2003}), spontaneous ocular fixation durations while participants were inspecting photographs (\emph{DOVES} database \cite{DOVES:2009}), and a sample of different web-collected experiments extracted from the \emph{PsychExperiments} \cite{McGraw:Tew:Williams:2000} web site. This last set included two-choice decision reaction times to both auditory (tones) and visual (colours) stimuli, reaction times of participants performing a mental rotation task, and the times that participants took to exit a virtual maze.

The solid dots in the left panel of Fig.~1 plot (in log-log scale) the histograms of the {RL}s in each of the datasets, aggregated across participants. Notice that all six distributions show a very similar pattern: The probabilities of the faster latencies rise to a peak, from which they decrease, gradually approaching a straight line, which is the characteristic signature of a power-law distribution. As predicted, the straight line components seem remarkably parallel across the six datasets, with a slope of approximately minus two (black dot-dashed lines). The right panel in Fig.~1 further stresses this apparent invariance. It plots the corresponding distributions when the times have been divided by their medians so as to remove the scale-dependent component of the distributions. One can distinguish three phases in the distributions. The early times rise to a peak, following very different patterns for each dataset. From the mode up to somewhere between five and forty times the median, there is a transition phase where the distributions gradually approach a power-law. The precise speed of convergence to the power-law varies depending on the properties of the participant and the task \cite{Moscoso:2009,Holden:VanOrden:Turvey:2009}. From this point onwards -- as stressed by the inset panel in the figure -- the distribution of latencies is approximately the same, regardless of the particular experimental task. To confirm that this pattern holds when one considers only single-participant data, the figures also plot the histogram of the responses of an individual participant in the Dundee dataset (open red circles; these correspond to participant ``sd'', but the pattern also holds for all other participants). Finally, in order to illustrate the theoretical prediction across the whole range of latencies, the figure also includes the theoretical density that would be predicted by an instance of the {NRD} with arbitrary parameters (black solid lines), and how the histogram from a sample of such would look like (grey open circles).

The histograms in Fig.~1 seem consistent with the hypothesis that the right tails of latency distributions follow a power-law with a scaling parameter of two, and most certainly discard the traditional assumption of a light, exponential-type tail. However, other heavy-tailed distributions could also produce histograms with this appearance, and this has given rise to disagreements with respect to the precise nature of heavy tails in some datasets. Therefore, the hypothesis needs to be contrasted with other possible distributions with similarly heavy tails. Both log-normal and stretched-exponential ({\em i.e.,} Weibull) tailed distributions also give rise to very heavy tails \cite{Limpert:Stahel:Abbt:2001,Mitzenmacher:2004,Clauset:Shalizi:Newman:2007}, and both have been proposed as plausible theoretical or empirical models for {RL} distributions \cite{Logan:1992,Luce:1986}. In addition, as I predict that the power-law should have a scaling parameter of exactly two, any other power-law with an arbitrary scaling parameter -- not necessarily, but also including two -- could be an alternative description \cite{Holden:VanOrden:Turvey:2009}.

Tab.~1 summarises the posterior evidence supporting the hypothesis that the right tails follow a power law distribution with a scaling parameter of exactly two over each of the other three candidate hypotheses \cite{Kass:Raftery:1995}. For four out of the six aggregated datasets, and for the individual participant analysis, the evidence supports the hypothesis over the three competing candidates ({\em i.e.}, positive values in the table). In the remaining two cases (negative values, highlighted in bold), the best candidate distribution was a power-law with an arbitrary value of the scaling parameter. In both of these cases, it seems like the optimal value of the scaling parameter estimated under a general power-law hypothesis has a value above two (last row in the table).

The model comparison method was particularly stringent on the target hypothesis. The implicit truncation (see Supplementary Materials) present in the data could lead to the over-estimation of the scaling parameter that was found for two of the datasets. To investigate this possibility, I fitted an {NRD} to the {RL}s in the Maze dataset, as this was the one for which the theory showed the worst performance. From the fitted distribution, I generated a sample of artificial {RL}s of the same size as the Maze dataset, sampling only points below 50 times the median (this is equivalent to an upper truncation at around eight minutes). The sample was discretised to simulate a measurement resolution of one ms. Fig.~2 compares the original data with the sample and the fitted distribution. Although these simulated data originated from a power-law tailed distribution with a true scaling parameter of exactly two, applying the hypothesis testing procedure revealed a very similar pattern to what was observed in the Maze data (see the last column of Tab.~1). All three alternative hypotheses seemed more probable than the target (and correct) hypothesis due to the advantage that truncation gives them. Given the quality of the fit in Fig.~2, it seems likely that the same distortion took place in the Maze data. In sum, all datasets were consistent with the theoretical prediction. The theory was the best of the four candidate theories for the majority of the datasets studied, all of which showed power-law tails. One cannot fully discard that the power-law tail may be subject to a cutoff at some unknown high value. However, lacking a specific value for the location of the cutoff is equivalent to stating that the power-law regime continues indefinitely, which is also the most parsimonious assumption by the Principle of Maximum Entropy \cite{Bercher:2008}.

Power-laws are often interpreted as evidence for Self-Organizing Criticality ({SOC} \cite{Bak:Paczuski:1995}), but several other mechanisms could also give rise to power-laws without the explicit need for self-organization \cite{Newman:2005,Sornette:2001}. In the domain of human {RL}s, some authors have argued for the presence of {SOC} using evidence for $1/f$ `pink' noise in the frequency spectra of {RL}s \cite{Gilden:Thornton:Mallon:1995,VanOrden:Holden:Turvey:2003,Thornton:Gilden:2005,VanOrden:Holden:Turvey:2005}, but this evidence is currently subject to discussion \cite{Wagenmakers:Farrell:Ratcliff:2005,Farrell:Wagenmakers:Ratcliff:2006}. The fixed scaling parameter of two is common to a prototypical model of a system that is known to be in a critical state: The Ising model of a magnet \cite{Jan:etal:1999}. This model is also described by an {NRD}, originating from the fractional change in magnetization ($\Delta m / m$). At a small environment around its critical temperature, the Ising model exhibits power-law behavior, but very small deviations from the critical temperature restrict the power-law to the very far tails \cite{Jan:etal:1999}. This is very much what I have observed in the {RL}s. The power-law behavior settles at the extreme right tails, between five and forty times the median {RL} in a particular task. Rather than evidence for {SOC}, the results in fact argue for a system that has been pushed slightly away from its critical point. This suggests that, at rest, the system is likely to be in a state which could be characterised as {SOC}, but the presentation of stimuli disturbs this criticality. This picture is consistent with recent work on electro-physiology. Human (and animal) neural oscillations are generally at a critical state, characterised by both power-laws and  $1/f$ noise patterns, but transient synchronization of neural assemblies during cognitive processing can temporarily disturb this criticality \cite{Buzsaki:Draguhn:2004}. The power-law with exponent two provides a characterisation of this critical state. Measuring the magnitudes of deviations from this resting state elicited by different conditions can provide a direct measure of the amounts of information processing they involve, considered here as a relaxation in return to the critical state. 

It comes as no surprise that human behavior, given its neural origins, should be best described by a complex system. It has recently been suggested that scale-invariance, as expressed by power-laws, may constitute a ``universal principle'' governing human cognition \cite{Chater:Brown:2008}, and biological systems in general \cite{West:Brown:2005}, a view that is supported by this study.

{\footnotesize
\section*{Methods}

For an objective rationale to choose among the four possible explanations for the heavy tails, I used pairwise Bayes factors \cite{Kass:Raftery:1995} between the log-likelihoods for each candidate hypothesis. The distribution proposed here has no free parameters, thus the computation of the log-likelihood for a fixed value of $t_\mathrm{min}$ is straightforward (see additional materials for details on the selection of $t_{\min}$). However, the other three candidate hypotheses have either one (for the general power-law case) or two free parameters (the log-normal and Weibull tail cases). For these, the log-likelihood was computed by numerical integration on the parameter space, assuming truncated uninformative ({\it i.e.}, Jeffreys') priors for the free parameters. The truncation was designed to favour the three alternative hypotheses over the power-law proposed here. The integration space was restricted to plausible values of the parameters: For the general power-law hypothesis, I assumed that the scaling parameter should take a value greater than one and smaller than six, as power-laws with scaling paramaters greater than this are seldom observed in natural phenomena \cite{Clauset:Shalizi:Newman:2007,Mitzenmacher:2004,Newman:2005}. For the Weibull hypothesis, I assumed that the value of its shape parameter shoud never be above one, as this would imply an exponential tail or lighter, which cannot correspond to the pattern observed in the histograms. Finally, both the Weibull location and the log-normal location and scale parameters were restricted to values that would make the datapoints correspond to an actual right tail ({\it i.e.}, their mode should fall to the left of $t_\mathrm{min}$) and have a peak within the range of {RL}s. Note that these constraints actually increase the likelihood of the alternative hypotheses beyond the under-specification than is found in the literature, thus providing conservative estimates of the evidence in support of our hypothesis. Further technical details are provided in the Supplementary Materials.
}


\begin{description}
 \item [Acknowledgements]This work was partially supported by the European Commission through a Marie Curie European Reintegration Grant ({MERG-CT-2007-46345}). The author wishes to thank F.-X. Alario, B. Burle, P. Milin, A. Rey, and J. Ziegler for discussion and suggestions on these ideas, S. Dufau for help in processing eye-movement data, and A. Kennedy and I. van~der~Linde for providing access to their eye-movement databases.
 \item[Competing Interests] The author declares that he has no
competing financial interests.
 \item[Correspondence] Correspondence and requests for materials
should be addressed to F. M. d. P.~(email: {\tt fermin.moscoso-del-prado@univ-provence.fr}).
\end{description}

\newpage


\begin{table}[h]
\caption{{\bf Posterior evidence in favour of a power-law with exponent two} over each of the alternative hypotheses for the datasets analyzed, as well as for the artificially generated data simulating the Maze {RL}s. Positive values indicate support for the power-law with $\alpha=2$, while negative values indicate evidence in favour of each of the alternative hypotheses. The first two rows indicate the value of $t_\mathrm{min}$ in relation to the corresponding median, and the number of points above this threshold found in each dataset. The last row indicates the posterior estimate of $\alpha$ if one assumed the general (unrestricted exponent) power-law hypothesis to be true.}
\label{tab:comparisons}
\resizebox{\textwidth}{!}{
\begin{tabular}{ccccccccc}
\\
\hline
\hline
& Dundee & \multirow{2}{*}{DOVES}	& \multirow{2}{*}{Tones} & \multirow{2}{*}{Colours} &  \multirow{2}{*}{Rotation}	& \multirow{2}{*}{Maze} & Dundee & Simulated \\
& (whole set) & & & & & & (participant ``sd'') & Data\\
\hline
$t_\mathrm{min}/\mathrm{median(t)}$ & 10 & 10 & 5 & 5 & 40 & 10 & 5 & 10 \\
Number of $t\, \geq \, t_{\mathrm{min}}$ & 33 & 57 & 133 & 544 & 31  & 458 & 27 & 366\\
\hline
Log-Normal	& 5.5 dB & 2 dB & -6.5 dB & 13.5 dB & 13 dB & -257.5 dB & 9 dB & -15 dB \\
\\
Weibull 		& 2 dB  & 2 dB & 31 dB & 43 dB & 26 dB & -241.5 dB & 11.5 dB & -7 dB\\
\\
Power-Law (general) 	& 2 dB & 3 dB & {\bf -22.5 dB} & 9.5 dB & 5.5 dB & {\bf -261 dB} & 5.5 dB & \bf{-21 dB}\\
$\hat{\alpha}$ & 2.56 & 1.82 & {\bf 2.50} & 1.92 & 2.27 & {\bf 2.99} & 2.17 & {\bf 2.48} \\
\hline
\hline
\end{tabular}}
\end{table}

\begin{figure}[h]
\begin{center}
\includegraphics[width=\textwidth]{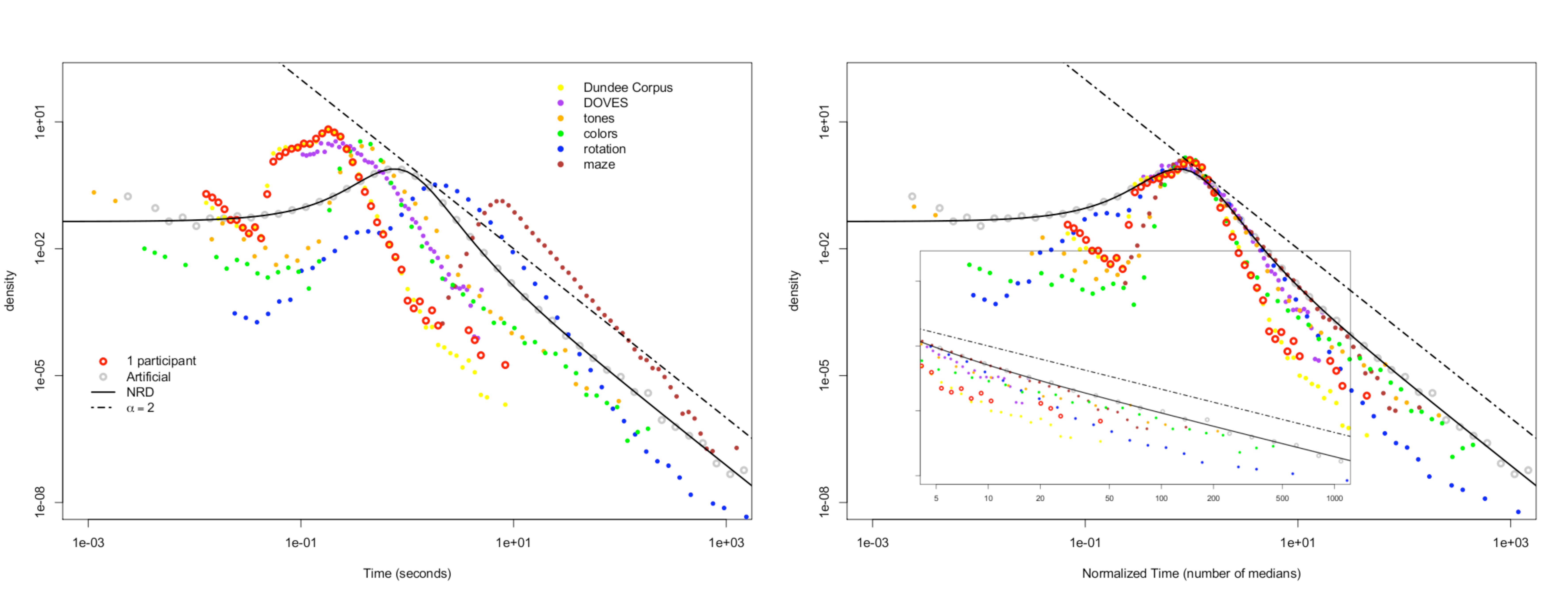}
\end{center}
\caption{Histograms of the latencies in second in the six datasets plotted in log-log scale. The solid dots represent the six datasets aggregated across participants. The open red circles plot the histogram from a single participant from the {\it Dundee} dataset. The open grey circles plot a sample from an arbitrary instance of the {NRD}, whose density corresponds to the solid black lines. The dot-dashed black lines illustrate a slope of -2. Both panels represent the same data either on the true time scale (left panel), or in the time scale normalised by the corresponding median (right panel). The inset on the right panel magnifies the power-law right tail of the distributions.}
\label{fig:aggregated}
\end{figure}

\begin{figure}[h]
\begin{center}
\includegraphics[width=\textwidth]{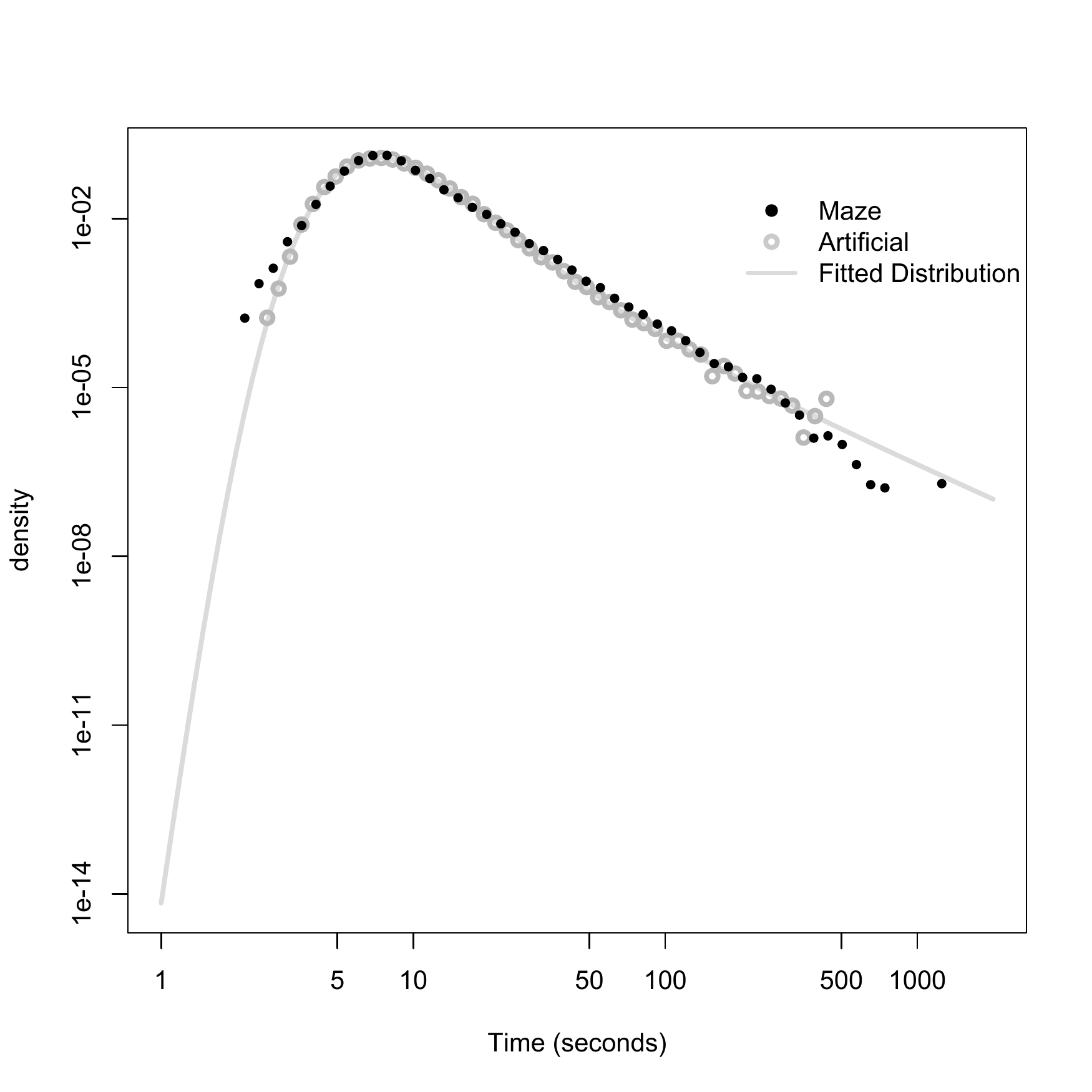}
\end{center}
\caption{Histograms (in log-log scale) of the {RL}s in the Maze dataset (black solid dots) and of the simulated artificial dataset (grey open circles). The solid grey line plots the Maximum Likelihood  NRD fit to the Maze dataset from which the simulated points were sampled.}
\label{fig:blinks}
\end{figure}

\clearpage

\appendix

\setcounter{table}{0}
\setcounter{equation}{0}
\setcounter{figure}{0}
\setcounter{page}{1}
\renewcommand{\thepage}{S\arabic{page}}
\renewcommand{\theequation}{S\arabic{equation}}
\renewcommand{\figurename}{Supplementary Figure}
\renewcommand{\tablename}{Supplementary Table}

\section*{Supplementary Materials}

\subsection*{Supplementary Methods}

\paragraph*{Materials}

The data from {\it PsychExperiments} correspond in their greater part to latencies of psychology students performing the experiments during in-class sessions, but some residual component of latencies originating from persons `trying out' the system is also present. In order to minimise in as a much as possible the distortion of the distributions introduced by users testing the system, only {RL}s from participants whose number of responses in the database corresponds exactly the number of stimuli in the experiment, were kept. In addition, all {RL}s with values equal or smaller than zero, or for which some field of the downloaded file was contained an incosistent value ({\it e.g.}, an incorrect task descriptor, {\it etc.}) were removed prior to the analyses.  Descriptive statistics of the resulting datasets are provided in Supl.~Tab.~1.

The experimental {RL}s were measured to a precision of 1 msec., except for the ocular {RL}s from {\it DOVES}, which had a precision of 5 msec. ({\em i.e.}, the eye-tracking equipment used a sampling rate of 200 Hz.). A similar discretization was perfomed on the artificially generated datasets, simulating an experimental resolution of 1 msec. relative to a median of 9,278 msec. estimated by maximum likelihood. 

\paragraph*{Histograms}

To obtain a smoother estimate of the far-right tails of the histograms I used logarithmically increasing bin widths for the histograms [21]. Note that this technique over-estimates the densities in the lower time ranges when the data have been discretised by the measurement apparatus.

\paragraph*{Estimation of the evidence}

For any two candidate models $M_1$ and $M_2$, if $\mathbf{t} = (t_1,\ldots,t_N)$ are the latencies in the dataset that are longer than $t_\mathrm{min}$, the evidence in favor of $M_1$ over $M_2$ is:
\begin{equation}
E(M_1,M_2|\mathbf{t}) = 10\log_{10} \frac{p(M_1|\mathbf{t})}{p(M_2|\mathbf{t})} = 10 \log_{10} \frac{p(M_1)}{p(M_2)} + 10 \log_{10} \frac{p(\mathbf{t}|M_1)}{p(\mathbf{t}|M_2)},
\end{equation}
where the second step is achieved by application of Bayes' Theorem, and $p(\mathbf{t}|M_i)$ is the likelihood of the datapoints for model $M_i$. If {\em a-priori} we consider both models equally probable, the evidence reduces to the difference in log-likelihoods:
\begin{equation}
E(M_1,M_2|\mathbf{t}) = 10 \left[ \log_{10} p(\mathbf{t}|M_1) - \log_{10} p(\mathbf{t}|M_2)\right] = 10 \left[ \ell(\mathbf{t}|M_1) - \ell(\mathbf{t}|M_2)\right], 
\end{equation}
I used decimal logarithms and the scaling factor of $10$ so that the resulting evidence would be measured in decibels [31].

\paragraph*{Log-likelihood for the target distribution} For the power-law with a pre-determined scaling parameter $\alpha=2$, the computation of $\ell$ is straightforward:
\begin{equation}
\ell(\mathbf{t}|\alpha=2) = \log_{10} \prod_{i=1}^N p(t_i|t_{\mathrm{min}},\alpha=2) = \sum_{i=1}^N \log_{10} p(t_i|t_{\mathrm{min}},\alpha=2),
\end{equation}
where $p(t_i|t_{\mathrm{min}},\alpha=2)$ is the density function of a discrete power-law (normalised for an upper truncation at $\mathrm{max}\{\mathbf{t}\}$).

\paragraph*{Log-likelihood for the alternative hypotheses} For an hypothesis $M$ with free parameters $\boldsymbol{\theta}=(\theta_1, \ldots, \theta_k)$, the marginal log-likelihood of the hypothesis is given by:
\begin{equation}
\ell(\mathbf{t}|M)  =  \log_{10} \int_{V(\boldsymbol{\theta})} p(\mathbf{t},\boldsymbol{\theta}|M) \mathrm{d}\boldsymbol{\theta} 
 =  \log_{10} \int_{V(\boldsymbol{\theta})} p(\mathbf{t} | \boldsymbol{\theta},M) p(\boldsymbol{\theta}|M) \mathrm{d}\boldsymbol{\theta},
\label{eq:integral}
\end{equation}
where $V(\boldsymbol{\theta})$ is the volume defined by the parameters, $p(\boldsymbol{\theta}|M)$ is the prior on $\boldsymbol{\theta}$ and:
\begin{equation}
p(\mathbf{t}|\boldsymbol{\theta},M) = \prod_{i=1}^{N} p(t_i|\boldsymbol{\theta},M),
\end{equation}
where $p(t_i|\boldsymbol{\theta},M)$ is given by the density function of $M$.

For each hypothesis, I estimated the integral in Eq. S4 numerically using a Montecarlo technique. I sampled $10^5$ points from the prior distribution $p(\boldsymbol{\theta}|M)$, computed the likelihood of $\mathbf{t}$ using the sampled values of $\boldsymbol{\theta}$, and took the mean result as the marginal likelihood. For each of the three distributions, I used discrete versions of their densities$^{16}$ truncated between $t_\mathrm{min}$ and $\mathrm{max}\{\mathbf{t}\}$.

In order to minimise numerical errors, all the computations above were performed in logarithmic scale. Evidence values were rounded to half-decibels, as this is argued to be minimal perceptible difference in evidence [31].

\paragraph*{Selection of $t_\mathrm{min}$}

The threshold value $t_\mathrm{min}$ was chosen by visual inspection of the histograms in Fig.~1. In addition, these choices were validated by assessing whether the chosen value would be close to the one that would minimise the value of the Kolmogorov-Smirnoff statistic between the sample of {RL}s and a general power-law [16], which in all cases suggested similar values. Note however, that this objective method can be problematic when one considers the additional upper truncation that is present in our data.

\begin{table}[h]
\caption{Description and summary statistics of the datasets used in the analyses.}
\label{tab:datasets}
\resizebox{\textwidth}{!}{
\begin{tabular}{ccccccccc}
\\
\hline
\hline
	  & 			&			& 				& & & & Reading\\
Task & Reading	& Visual 		& Two-choice {RT}	& Two-choice {RT} & Mental & Maze & (single & Artificial\\
	  & 			& inspection	& 	& & rotation & &  subject) & (truncated)\\
\hline
\hline
\multirow{2}{*}{Stimulus} 	& visual & visual	& auditory	& visual	& visual	& visual   & visual & \multirow{2}{*}{--} \\
					& dynamic & static	& static	& static	& static	& dynamic & dynamic \\ 
\hline
\multirow{2}{*}{Response}	& ocular	& ocular	& manual	& manual	& manual	& manual & ocular & \multirow{2}{*}{--} \\
						& simple	& simple	& simple	& simple	& simple	& complex & simple \\
\hline
N. responses	& 420,809 & 35,500 & 17,800 & 107,360 & 589,440 & 63,090 & 50,731 & 63,090 \\
N. participants	& 10 & 29	& 890 & 5,368 & 12,280 & 4,206 & 1 & --  \\
Resp./Part. & 42,080.90 & 1,224.13 & 20.00 & 20.00 & 48.00 & 15.00 & 50,731 &--  \\
Minimum  & 0.012 s & 0.100 s & 0.010 s		& 0.030 s		& 0.021 s & 2.007 s & 0.012 s &  2.528 s\\
Maximum & 8.901 s & 5.200 s & 108.657 s	& 172.211 s	& 3,561.324 s & 1,331.391 s & 8.901 s & 460.460 s\\
Mean $t$ & 0.195 s & 0.372 s & 0.590 s & 0.493 s & 3.478 s & 13.101 s   & .195 s &  12.483\\
Median $t$ & 0.188 s & 0.295 s & 0.661 s & 0.416 s & 2.692 s & 9.019 s &  .188 s & 9.273 s \\
\hline
\hline
\end{tabular}}
\end{table}

\subsection*{Supplementary Notes}

\begin{itemize}
\item {\bf Precision of on-line {RL} measurements}  The datasets from {\em PsychExperiments} were collected online using authorware programs. In general, the precision of these measurements is approximately equivalent to the precisions obtained in traditional laboratory testing sessions ($\sim$~1~ms.), except for cases in which the client computer has an exceptionally large number of concurrent processes running [13,32]. 
\item {\bf Implicit truncation} An additional problem that affects the comparison between the different theories is the right-truncation that is present in the data. This introduces a bias in favor of lighter tails: It favors either of the non-power-law distributions, or the general power-law with a high value of the scaling parameter. In both of the eye movement datasets, fixations are determined by computer algorithms that rely on aspects such as their duration, therefore implicitly introducing both left and right truncations. In the internet collected datasets, there is officially no right truncation in the data. However, considering that the great majority of the data were collected during in-class training sessions, most responses will be subject to an implicit upper bound dictated by the duration of the class. The residual of responses longer than one class session are very likely to originate in users testing the system, system failures in the client, {\it etc.}. To attenuate this problem, I assumed that all datasets had been right-truncated at the maximum {RL} observed. For the web collected datasets this is still an under-estimation of the real truncation point, leaving some advantage for the non-power-law distributions, and over-estimating the scaling parameter for power-laws. This problem is more acute in the longer {RL} datasets, where the truncation point comes closer to the magnitudes of valid observed {RL}s.
\item {\bf Supplementary references} 
\begin{enumerate} 
\setcounter{enumi}{30}
\item E. T. Jaynes, {\it Probability Theory: {T}he Logic of Science} ({Cambridge University Press}, Cambridge, {UK}, 2003).
\item M. D. Tew, K. O. McGraw, {\em The Accuracy of Response Timing by Authorware Programs} (Unpublished Manuscript, University of Mississippi, 2002).\\  Available from {\tt http://psychexps.olemiss.edu/Scrapbook/Timing\_2002.pdf}.
\end{enumerate}
\end{itemize}

\end{document}